\documentclass[12pt]{article}

\usepackage{amssymb,epsfig,cite}

\newcommand{\be}{\begin{equation}}
\newcommand{\ee}{\end{equation}}

\newcommand{\ri}{{\rm i}}
\newcommand{\ra}{\rangle}
\newcommand{\la}{\langle}
\newcommand{\CC}{\mathbb{C}}
\newcommand{\C}{{\kern+.25em\sf{C}\kern-.48em\sf{I} \kern+.48em\kern-.25em}}

\begin{document}

\begin{center}
{\large\bf NEUTRINOS: \vspace*{2mm} \\
Mysterious Particles with Fascinating Features,
\vspace*{3mm} \\ 
which led to the Physics Nobel Prize 2015}

\vspace*{5mm}

Alexis Aguilar-Arevalo and Wolfgang Bietenholz \vspace*{2mm} \\
Instituto de Ciencias Nucleares \\
Universidad Nacional Aut\'{o}noma de M\'{e}xico \\
A.P.\ 70-543, C.P.\ 04510 Distrito Federal, Mexico
\end{center}

\vspace*{3mm}

\noindent
{\em The most abundant particles in the Universe are photons
and neutrinos. Both types of particles are whirling around
everywhere, since the early Universe. Hence the neutrinos 
are all around us, and permanently pass through our planet and our 
bodies, but we do not notice: they are extremely elusive. 
They were suggested as a theoretical hypothesis in 1930,
and discovered experimentally in 1956. 
Ever since their properties keep on surprising us; for
instance, they are key players in the violation of parity symmetry.
In the Standard Model of particle physics they appear in three
types, known as ``flavors'', and since 1998/9 we know that 
they keep on transmuting among these flavors. This
``neutrino oscillation'' implies
that they are massive, contrary to the previous picture,
with far-reaching consequences. This discovery
was awarded the Physics Nobel Prize 2015.}

\section{A desperate remedy}

ETH Z\"{u}rich, the Swiss Federal Institute of Technology, has a long
tradition of excellence in physics and other sciences. In addition, 
it has a tradition (dating back to the 19th century) to celebrate
each year a large dance event,
the Polyball. This also happened in 1930, when Wolfgang Pauli, one of 
the most renowned theoretical physicists, was working at ETH.
The Polyball prevented him from attending a workshop in T\"{u}bingen
(Germany), where leading scientists met to discuss aspects
of radioactivity. Instead Pauli sent a letter to the 
participants, whom he addressed as ``Liebe Radioaktive Damen 
und Herren'' (``Dear Radioactive Ladies and Gentlemen'') \cite{Pauli}. 
This letter of one page was of groundbreaking importance: it was
the first document where a new type of particle was suggested, 
which we now denote as the {\em neutrino}. 

Pauli was referring to the
energy spectrum of electrons emitted in the $\beta$-decay:
from a modern perspective (not known in 1930), a neutron
is transformed into a slightly lighter proton, emitting an electron. 
This {\em $\beta$-radiation} was observed, but the puzzling
point was the following: there is some energy reduction in a
nucleus where this decay happens, and if we subtract the 
electron mass, we should obtain the electron's kinetic energy, which 
ought to be the same for all electrons emitted.
In fact, the $\alpha$- and $\gamma$-radiation spectra do
exhibit such a sharp peak. For the $\beta$-radiation,
however, one observed instead a broad spectrum of electron 
energies \cite{Chadwick14}, with a maximum at this value. 
In particular, in 1927 C.D.\ Ellis and W.A. Wooster had
studied the decay $^{210}_{83} {\rm Bi} \to \ ^{210}_{84} {\rm Po}$
and identified a maximal electron energy of 1050 keV, but a
mean value of only 390 keV \cite{EllisWooster}.
\begin{figure}
\begin{center}
\includegraphics[angle=0,width=.4\linewidth]{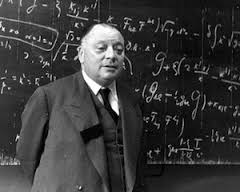}
\hspace*{8mm}
\includegraphics[angle=0,width=.43\linewidth]{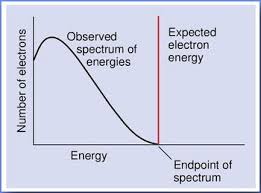}
\end{center}
\caption{\it{On the left: Wolfgang Pauli (1900-1958),
Austrian physicist working in Z\"{u}rich, Switzerland.
On the right: the energy spectrum of the electron, which
is emitted in a $\beta$-decay; the observation does not
match the original expectation of a sharp peak. Pauli solved this 
puzzle by postulating the emission of an additional particle,
which was hypothetical at that time.}}
\vspace*{-3mm}
\end{figure} 

This seemed confusing indeed, and prominent people like Niels Bohr even 
considered giving up the law of energy conservation. Pauli, however, 
made an effort to save it: as a ``desperate remedy'' he postulated
that yet another particle could be emitted in this decay, which
would carry away the energy, which seemed to be missing. He estimated 
its mass to be of the same order as the electron mass. He also
knew that some nuclei change their spin by 1 unit under $\beta$-decay,
so he specified that this new particle should carry spin 1/2,
just like the electron; thus also angular momentum conservation
is saved. To further conserve the electric charge, it must be
electrically neutral, therefore he wanted to call it a ``neutron''.
That would explain why this particle had not been observed, thus
completing a hypothetical but consistent picture.\footnote{Hence
Pauli suggested one new particle, for truly compelling reasons 
like the conservation of energy and angular momentum. This can be 
contrasted with the modern literature, where a plethora of hypothetical
particles are suggested, often based on rather weak arguments.}

\section{Fermi's theory}

Two years later, James Chadwick discovered the far more massive particle, 
which we now call the neutron \cite{neutron}. In 1933/4 Enrico Fermi, 
who was working in Rome, elaborated a theory for the interaction 
of Pauli's elusive particle \cite{Fermi}. He introduced 
the name ``neutrino'',\footnote{Since ``neutrino'' is a diminutive
in Italian, its plural should actually be ``neutrini'', but we
adopt here the commonly used plural.}
and suggested that it might be {\em massless.}
\begin{figure}
\begin{center}
\includegraphics[angle=0,width=.4\linewidth]{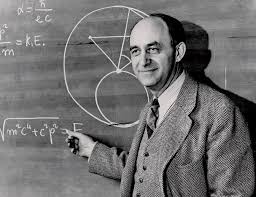}
\hspace*{8mm}
\includegraphics[angle=0,width=.4\linewidth]{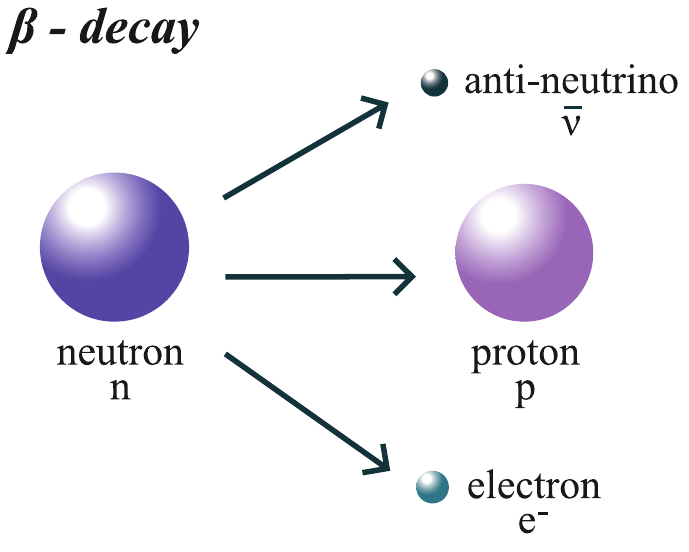}
\end{center}
\vspace*{-5mm}
\caption{\it{On the left: Enrico Fermi (1901-1954), famous 
for his achievements both in theoretical and experimental physics.
On the right: scheme of the $\beta$-decay, which transforms a neutron
into a proton, while emitting an electron and an anti-neutrino.}}
\end{figure} 

In our modern terminology, the emitted particle is actually an
{\em anti-neutrino,} $\bar \nu$. This $\bar \nu$-emission is,
in some sense, equivalent
to an incoming neutrino, $\nu$, so the $\beta$-decay can be written
in its usual scheme, or as a related variant,
$$
{\rm n} \to {\rm p} + {\rm e}^{-} + \bar \nu 
\qquad {\rm or} \qquad
{\rm n} + \nu \to {\rm p} + {\rm e}^{-} \ .
$$
Referring to the latter scheme, Fermi made an ansatz for the
transition amplitude $M$, where the wave functions of all four fermions 
interact in one space-time point $x$ (to be integrated over),
\be \label{4fermi}
M(x) = G_{\rm F} \, \Big( \bar \Psi_{\rm p}(x) \Gamma \Psi_{\rm n} (x) 
\Big) \, 
\Big( \bar \Psi_{\rm e}(x) \Gamma' \Psi_{\nu} (x) \Big) \ , \quad
G_{\rm F} \simeq 1.2 \cdot 10^{-5} \frac{(\hbar c)^{3}}{{\rm GeV}^{2}} \ . \
\ee
This {\em 4-fermi term} describes the simultaneous transformations
${\rm n} \to {\rm p}$ and $\nu \to {\rm e}^{-}$, with factors
$G_{\rm F}$ (Fermi's constant),\footnote{It is remarkable that
Fermi already estimated its magnitude correctly, his value 
was $G_{\rm F} = 0.3 \cdot 10^{-5} (\hbar c)^{3}/{\rm GeV}^{2}$.}
and $\Gamma, \ \Gamma '$ (to be addressed below).
In Heisenberg's formalism, these are just transitions between
the two isospin states of the same particle.\footnote{The nucleons,
{\it i.e.}\ the proton and the neutron, were assumed to be
elementary particles at that time.}

This process is a prototype of the {\em weak interaction}, 
which is nowadays described by the exchange
of $W$- and $Z$-bosons (Fermi's constant can be expressed as
$G_{\rm F} = g^{2}/ (2^{5/2} M_{W})$, where $g$ is the weak
coupling constant and $M_{W}$ the $W$-mass). 
Fermi's simple theory works well up to moderate
energy. The refined picture --- with an intermediate $W$-boson
instead of the 4-fermi interaction in one point --- 
prevents a divergent cross-section at high energy.

\section{Neutrinos exist!}

Pauli is often quoted as saying ``I have done a terrible thing,
I have postulated a particle that cannot be detected''
(although it is not clear where this statement is really
documented). In any case, it turned out to be wrong:
in 1956 Clyde Cowan and Frederick Reines observed that
anti-neutrinos, produced in a nuclear reactor in South Carolina,
did occasionally interact with protons, which leads to a neutron
and a positron (the positively charged anti-particle of an 
electron), ${\rm p} + \bar \nu \to {\rm n} + e^{+}$.
This is an {\em inverse $\beta$-decay,} which they observed
in two large water tanks \cite{CowRei}.\footnote{Even today, reactor 
neutrinos are still detected with a variant of the technique employed by 
Cowan, Reines and collaborators.} They sent a telegram to Pauli, alerting
him that his particle really exists!

\begin{figure}
\begin{center}
\includegraphics[angle=0,width=.4\linewidth]{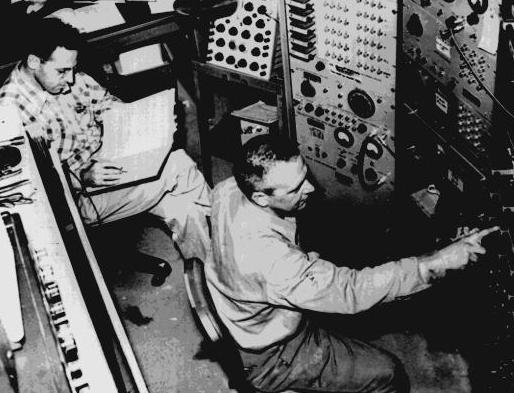}
\hspace*{5mm}
\includegraphics[angle=0,width=.5\linewidth]{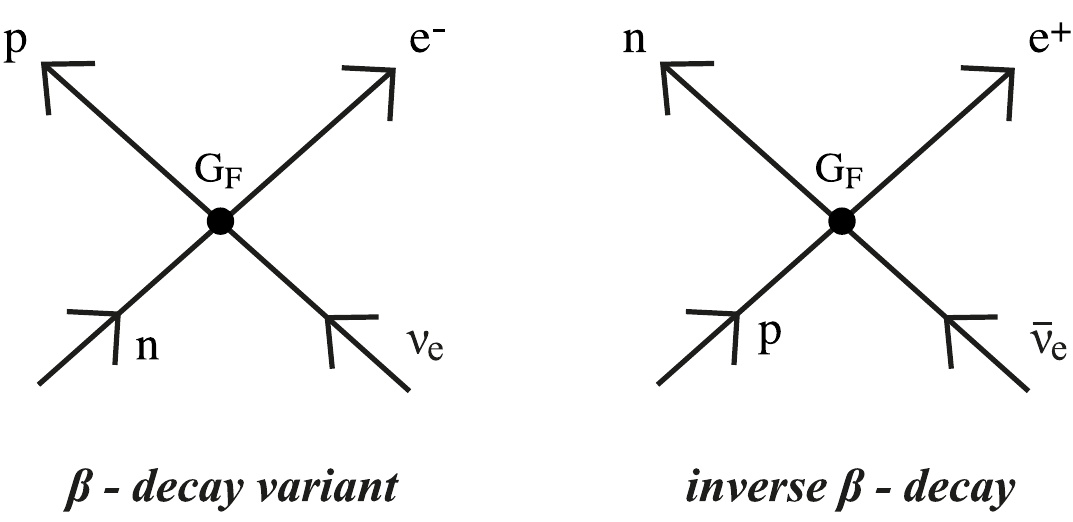}
\end{center}
\vspace*{-3mm}
\caption{\it{On the left: Fred Reines (left) and Clyde Cowan (right), 
the pioneers who first succeeded in detecting anti-neutrinos.
On the right: diagrams of a $\beta$-decay variant (compatible with 
Fermi's formula (\ref{4fermi})), and of the inverse $\beta$-decay
(observed by Cowan and Reines in 1956).}}
\vspace*{-2mm}
\end{figure} 

\subsection{$\dots$ and they are all around!}

Of course, neutrinos had existed long before, since the Big Bang:
just 2 seconds later they decoupled and ever since
they are flying around all over the Universe. This is the 
{\em Cosmic Neutrino Background,} C$\nu$B. It has gradually 
cooled down, from $\approx 10^{10}~{\rm K}$
to its temperature today of $1.95~{\rm K}$.
It can be compared to the (better known) Cosmic Microwave Background, 
which was formed about 380\,000 years later by photons,
and which is somewhat warmer, 2.73 K.

In contrast to the Cosmic Microwave Background, which is being
monitored intensively, the C$\nu$B has not been observed directly ---
neutrino detection is very difficult in general, and at such low energies 
it seems hardly possible. Still, the arguments for its existence
are compelling and generally accepted. New indirect evidence has 
been provided in 2015 by Planck satellite data for details
of the temperature fluctuations in the Cosmic Microwave Background 
\cite{Planck2015}. A {\em direct} detection of the C$\nu$B, however, 
is still a long-term challenge. The density throughout the
Universe is about 336 neutrinos (and 411 photons) per cm$^{3}$; 
in our galaxy it might be higher due to gravitational effects.

Neutrinos of higher energies are generated in stars 
--- like the Sun --- by nuclear fusion, in Active Galactic Nuclei, 
Gamma Ray Bursts, supernova explosions, etc. 
They are also produced inside the Earth (by decays), in our atmosphere 
(when cosmic rays hit it and trigger an air shower of secondary 
particles), and on the Earth, in particular in nuclear reactors. The latter
provide $\bar \nu$-energies around 1 MeV, with a 
typical cross section of about $10^{-44}~{\rm cm}^{2}$. 
The probability of an interaction in a solid detector of 1 m length 
is of order $10^{-18}$, so their chance of scattering while 
crossing the Earth is around $10^{-11}$.

This shows why it took a while to discover them; 
the search for neutrinos is sometimes described as ``ghost hunting''.
For instance, in our daily life we never feel that we are 
exposed to a neutrino flux originating from the Sun, although some
$6 \cdot 10^{14}$ solar neutrinos cross our body every second. 
If we could install a detector that fills all the space between the
Sun and the Earth, it would capture only 1 out of 10 million neutrinos. 
In Section 7 we will come back to the solar and atmospheric neutrinos; 
this is what the 2015 Nobel Prize experiments were about.

\section{Parity violation: a stunning surprise}

\subsection{Theory}

A parity transformation, P, is simply a sign change of the spatial
coordinates, P: $x = (t, \vec r) \to (t, - \vec r )$. For a long
time, people assumed it to a basic principle that the Laws of
Nature are parity invariant. This seems obvious by common sense,
and in fact it holds for gravity, electromagnetic and strong
interactions. How about the weak interaction?
The neutrinos are the only particles that only interact weakly
(if we neglect gravity), so it is promising to focus on them to 
investigate this question.

At this point, we come back to the factors $\Gamma$ and $\Gamma'$
between the fermionic 4-component Dirac spinors $\bar \Psi$, $\Psi$ 
in eq.\ (\ref{4fermi}).
They characterize the structure of the weak interaction, which
arranges for these particle transformations. {\it A priori} one could 
imagine any Dirac structure: scalar, pseudo-scalar, vector, pseudo-vector
or tensor ($1 \!\!1, \, \gamma_{5}, \, \gamma^{\mu}, 
\gamma^{\mu} \gamma_{5}, \sigma^{\mu \nu}$).
Under parity transformation, the ``pseudo''-quantities (which 
involve a factor $\gamma_{5}$) pick up a sign opposite to
(ordinary) scalar and vector terms.
 
If $\Gamma$ and $\Gamma'$ were both parity even, or both parity
odd, then also this weak interaction process would be parity 
symmetric. However, in 1956 Tsung-Dao Lee and Chen-Ning Yang 
suggested that this might not be the case \cite{LeeYang}.
Their scenario is reflected by a structure of the form
\be  \label{LeeYang}
M(x) = \frac{G_{\rm F}}{\sqrt{2}} \ \Big( \bar \Psi_{\rm p}(x) 
\gamma^{\mu}(1 -\frac{g_{\rm A}}{g_{\rm V}} \gamma_{5}) \Psi_{\rm n}(x) \Big) \,  
\Big( \bar \Psi_{\rm e}(x) \gamma_{\mu}(1 - \gamma_{5}) \Psi_{\nu}(x) \Big) \ , 
\ee
which mixes vector currents  --- which Fermi had in mind --- 
with pseudo-vector (or axial) currents.
The ratio $g_{\rm A}/g_{\rm V}$ is a constant;
its value is now determined as $\simeq 1.26$. 
Hence vector and axial vector currents are strongly mixed,
which breaks P invariance.
But how was the violation of parity symmetry verified?

\subsection{Experiment}

\begin{figure}
\begin{center}
\includegraphics[angle=0,width=.36\linewidth]{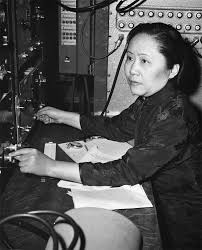}
\hspace*{10mm}
\includegraphics[angle=0,width=.36\linewidth]{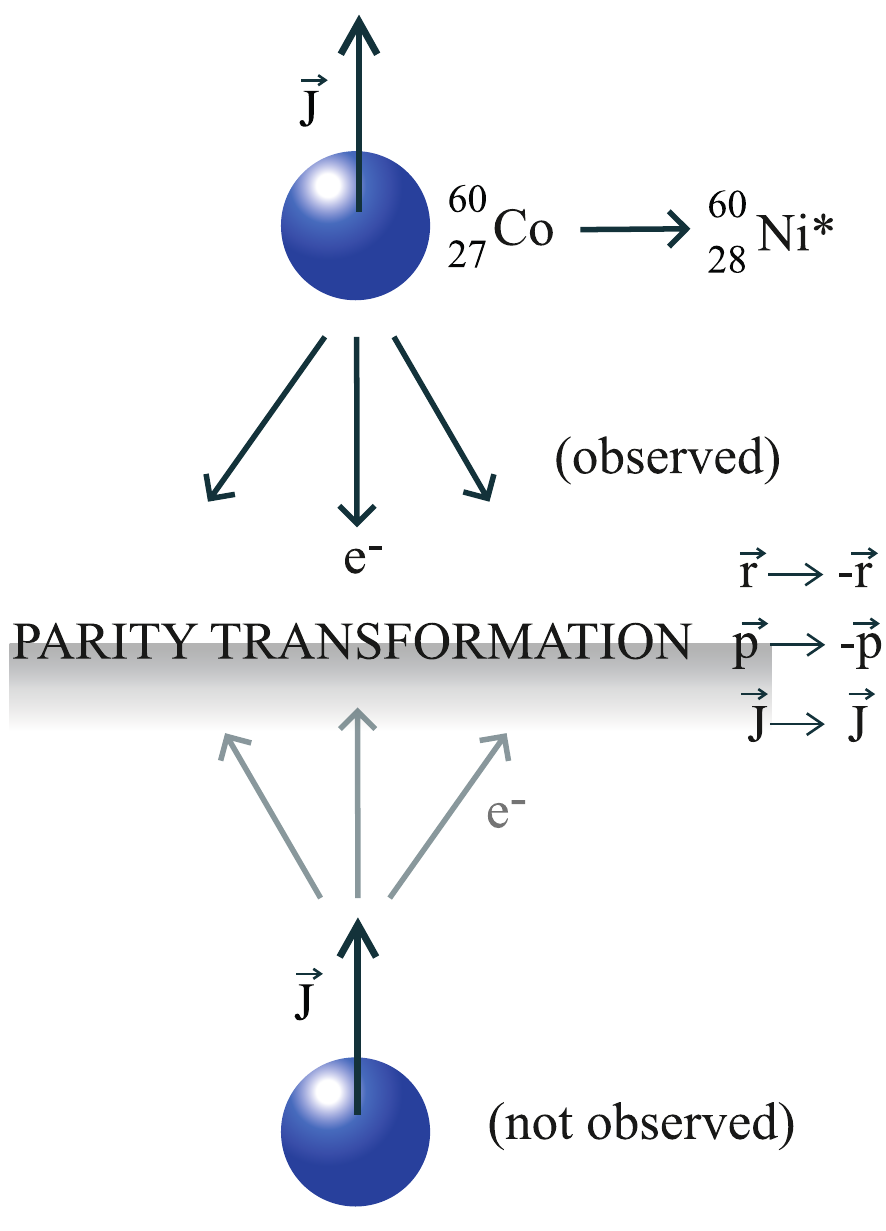}
\end{center}
\vspace*{-5mm}
\caption{\it{On the left: Chien-Shiung Wu (1912-1997), leader of the
experiment that demonstrated the violation of parity invariance in 1957.
On the right: the concept of her experiment, as described in the text.}}
\end{figure} 
In fact, it was confirmed only one year after Lee and Yang's
suggestion in an experiment, which was led by another brilliant Chinese 
researcher, Chien-Shiung Wu \cite{Wu}. Her experiment dealt with the
$\beta$-decay, which transforms a cobalt nucleus into nickel,
$$
^{60}_{27}{\rm Co} \ \to \ ^{60}_{28}{\rm Ni}^{*} + {\rm e}^{-} + \bar \nu \ ,
$$
a process, which lowers the nuclear spin from $J=5 \to 4$.
A magnetic moment is attached to the nuclear spin, hence
a strong magnetic field can align the spins in a set of Co nuclei. 
(This was not easy in practice: only after cooling the sample down 
to 0.003 K, a polarization of 60\% could be attained.)

How can the nuclear spin change be compensated by the leptons, {\it i.e.}\ 
by the electron e$^{-}$ and the anti-neutrino $\bar \nu \,$?
They are both spin-1/2 particles, as Pauli had predicted, and
they could be right-handed (spin in the direction of motion)
or left-handed (spin opposite to the direction of 
motion).\footnote{Strictly speaking this is the helicity, which 
coincides with the handedness, or chirality, in the relativistic 
limit; we are a bit sloppy about this distinction.}
Clearly, the compensation requires a right-handed particle 
flying away in the direction of the nuclear spin $\vec J$,
and a left-handed one being emitted in the opposite direction.

The electrons are much easier to detect, and one observed their
preference in the $-\vec J$ direction. 
Under a parity transformation, the spin $\vec J$ behaves like an angular 
momentum $\vec L = \vec p \times \vec r \,$; it remains invariant.
The direction of flight of the leptons, however, is exchanged.
Hence this dominance of electrons in one direction demonstrates the 
violation of parity invariance. The reason is that the anti-neutrino
only occurs right-handed (and the neutrino only 
left-handed),\footnote{\label{fnchiral} This can be seen from 
eq.\ (\ref{LeeYang}), which includes a projection of $\Psi_{\nu}$ 
to its left-handed component, 
$\psi_{\nu ;L} = {\textstyle\frac{1}{2}} (1 - \gamma_{5})\Psi_{\nu}$,
but no right-handed component
$\psi_{\nu ;R} = {\textstyle\frac{1}{2}} (1 + \gamma_{5})\Psi_{\nu}$
is involved. In fact, 
a right-handed neutrino, or a left-handed anti-neutrino, has never been 
observed. We will comment on their possible existence in the appendix.}
so the $\bar \nu$ has to move in the $\vec J$-direction.

This came a great surprise, {\em Nature does distinguish between
left and right!} An example for the consternation that this
result caused is Pauli's first reaction, who
exclaimed ``This is total nonsense!''.
It is a striking example for the fascinating 
features of the neutrinos. This sequence of surprises is still 
going on, and it embraces the 2015 Nobel Prize.
Long before, in 1957 Lee and Yang received the Nobel
Prize for their discovery; unfortunately Wu was left out.

As a {\it Gedankenexperiment}, 
one could also perform a C transformation
(``charge conjugation''), which transforms all particles into
their anti-particles and vice versa, thus flipping the signs
of all charges. This shows that the Wu
experiment also demonstrated the violation of C symmetry,
but invariance is recovered under the combined transformation CP.
In particular for the {\em chirality} (handedness) of $\nu$ and 
$\bar \nu$, CP invariance holds. Lev Landau suggested that
this might be a true symmetry of Nature \cite{Landau}.

In 1964, however, an experiment directed by James Cronin 
and Val Fitch demonstrated that --- in even more subtle decays, also
due to the weak interaction --- CP symmetry is violated as well. 
Now we are left with the CPT Theorem:\footnote{A rigorous proof for 
this theorem was given in 1957 by Res Jost \cite{Jost}, previously 
Pauli's assistant. It is one of the most important and elegant results 
in Quantum Field Theory, but it is not easily accessible: Jost wrote
his paper in German and published it in the Swiss journal {\it Helvetica 
Physica Acta,} which does not exist anymore.}
if we still add a simultaneous T transformation (a flip of the
direction of time), then invariance must hold, if our world
is described by a relativistic and local quantum field theory 
--- that seems to be the case, so far a huge number of high
precision experiments support it.

\section{Neutrinos occur in distinct flavors}

What distinguishes a neutrino from an anti-neutrino?
We have mentioned the different chirality.
In the Standard Model --- to be addressed below ---  left-handed 
neutrinos $\nu_{L}$ (right-handed anti-neutrinos $\bar \nu_{R}$) occur, 
and they carry a weak hypercharge $Y$ ($-Y$), 
which characterizes their coupling to a $W$ or $Z$
gauge boson (like the electric charge of other particles
represents the coupling to a photon). 
Thus also the sign of $Y$ distinguishes $\nu$ from $\bar \nu$.
However, their distinction was introduced much earlier,
even before either of them had been detected.

In 1953, E.J.\ Konopinski and H.M.\ Mahmoud studied the decays 
involving the light particles that we call leptons \cite{KonoMah}.
At that time, they knew the electron, the neutrino (as a hypothesis) 
and the muon, $\mu^{-}$, which had been discovered in 1936. The 
latter is similar to an electron, but 207 times heavier. 
Konopinski and Mahmoud introduced a new quantum number: they
assigned to the particles $\nu,\, e^{-},\, \mu^{-}$ the 
{\em lepton number} $L=1$, their anti-particles 
$\bar \nu,\, e^{+},\, \mu^{+}$ carry $L=-1$, and all the
(non-leptonic) rest has $L=0$.

The role of the lepton number should simply be its conservation,
which holds indeed {\it e.g.}\ in the $\beta$-decay, or
inverse $\beta$-decay, or in decays of charged pions,
\be  \label{pitomu}
\pi^{-} \to \mu^{-} + \bar \nu \ , \quad \pi^{+} \to \mu^{+} + \nu \ ,
\ee
but it rules out a process like 
${\rm n} + \bar \nu \to {\rm p} + {\rm e}^{-}$, 
which is not observed.

This rule is still incomplete, however, since it allows for
a decay like $\mu^{-} \to {\rm e}^{-} + \gamma$ ($\gamma$ 
represents a photon), which is not observed either.

This led to the insight that leptons occur in distinct 
{\em generations,} with their own lepton numbers, like the
electron number $L_{\rm e} = \pm 1$ for ${\rm e}^{\mp}$, and the muon 
number $L_{\mu} = \pm 1$ for $\mu^{\mp}$. This suggested that there are
also distinct neutrinos, as Bruno Pontecorvo --- an Italian physicist who 
had emigrated to the Soviet Union --- pointed out in 1960:
an electron-neutrino $\nu_{\rm e}$
with $L_{\rm e} = 1$ and a muon-neutrino $\nu_{\mu}$ with
$L_{\mu} = 1$ (while $\bar \nu_{\rm e}, \bar \nu_{\mu}$ have
$L_{\rm e} = -1$ and $L_{\mu} = -1$, respectively, and
the rest is zero) \cite{Pont60}. The stronger assumption
that $L_{\rm e}$ and $L_{\mu}$ are {\em separately conserved}
explains observed decays such as
\be  \label{mudecay}
\mu^{-} \to {\rm e}^{-} + \nu_{\mu} + \bar \nu_{\rm e} \ , \quad
\mu^{+} \to {\rm e}^{+} + \bar \nu_{\mu} + \nu_{\rm e} \ , 
\ee
which take $2.2 \cdot 10^{-6}$ s.
It also distinguishes transitions like
\be  \label{CC}
{\rm n} + \nu_{\rm e} \to p + {\rm e}^{-} \ , \quad
{\rm n} + \nu_{\mu} \to p + \mu^{-} \ ,
\ee
which require an intermediate charged boson $W^{\pm}$.
These transitions do not occur if we exchange $\nu_{\rm e}$ and
$\nu_{\mu}$, or replace them by anti-neutrinos. This distinction
enabled the experimental discovery of $\nu_{\mu}$ in 1962, 
by Lederman, Schwartz, Steinberger and collaborators \cite{LSS62}.
Now we can write the inverse $\beta$-decay, observed by Cowan 
and Reines, in a more precise form: 
${\rm p} + \bar \nu_{\rm e} \to {\rm n} + {\rm e}^{+}$.\\

The {\em Standard Model of particle physics} takes into account that 
later (in 1975) yet another cousin of the electron was found \cite{tau}:
the tauon $\tau$, which is 3477 times heavier than the electron
(hence its life time is only $2.9 \cdot 10^{-13}$ s).
It is also accompanied by its own type of neutrino \cite{nutau},
$\nu_{\tau}$, so we are actually dealing with {\em three} distinct
lepton numbers, $L_{\rm e}$, $L_{\mu}$ and $L_{\tau}$.

\begin{figure}
\begin{center}
\includegraphics[angle=0,width=.5\linewidth]{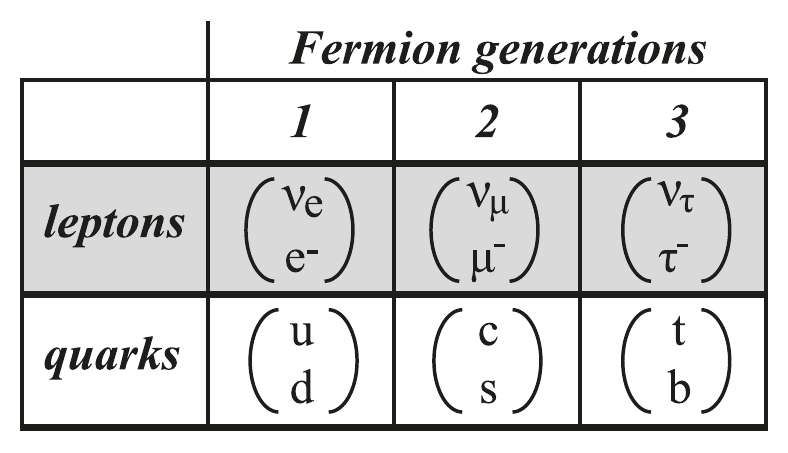}
\end{center}
\label{fermiontab}
\vspace*{-7mm}
\caption{\it{Table of the fermions in the Standard Model.}}
\end{figure}

Similarly the Standard Model incorporates three generations of 
quarks, so its fermionic content can be summarized as shown in
Table \ref{fermiontab}. 
In addition, the Standard Model involves gauge bosons (photons for 
the electromagnetic interaction, $W$ and $Z$ for the weak interaction,
and 8 gluons for the strong interaction), plus the (scalar) Higgs 
particle. This is what all known matter in the Universe consists 
of.\footnote{The graviton might still be added to this list.
We also have indirect evidence for Dark Matter, which must be 
of a different kind.}

From a conceptual point of view, the Standard Model is only 
consistent for entire {\em fermion generations,} composed of
a lepton doublet and a quark doublet (otherwise quantum effects
break gauge invariance \cite{ABJ}). On the other hand, there
is no theoretical constraint on the number of generations.
The higher generations involve heavier fermions,
so they were discovered later. Hence one could wonder if this 
sequence is going on, and
further generations will be discovered step by step.

This cannot be rigorously excluded, but there are good reasons to 
assume that there are not more than these 3 generations.
The $Z$-boson is one of the heaviest elementary particles 
that we know, with a mass of 91 GeV, and it can decay into
a neutrino--anti-neutrino pair of the same flavor,
$$
Z \to \nu_{x} + \bar \nu_{x} \ , 
\quad x \in \{ {\rm e}, \, \mu , \, \tau \} \ .
$$
It can also decay into ${\rm e}^{-} + {\rm e}^{+}$, $\mu^{-} + \mu^{+}$
or $\tau^{-} + \tau^{+}$, or into a quark--anti-quark pair. If we sum 
up all these decay channels (which were measured very precisely in 
the Large Electron-Positron Collider at CERN \cite{LEP}), we obtain 
--- to a good precision --- the full decay rate of the $Z$-boson.
This is an argument against a 4th generation: if the $Z$-boson
could decay into yet another $\nu$--$\bar \nu$ pair, we should have
noticed the missing part in this sum of decay channels.\footnote{A 
loophole in this argument are neutrinos, with a very heavy
mass $> m_{Z} /2 \simeq 46~{\rm GeV}$, which are, however, 
considered unlikely.}

\section{The mixing of quark and of lepton flavors}

\subsection{A look at the quark sector}

We follow the historical evolution and first discuss mixing
in the quark sector: we saw that
the quarks occur in 6 {\em flavors,} such as the ``strange'' $s$
quark. Also here quantum numbers were introduced, which indicate 
the quark contents of a specific flavor. For instance, the 
{\em strangeness} of a hadron\footnote{Hadrons are observable
particles, composed of quarks and gluons. One distinguishes
baryons (with 3 valence quarks, $(qqq)$) and mesons (with a
valence quark--anti-quark pair, $(q \bar q)$).}
counts the number of its $\bar s$ minus $s$ (valence) quarks.

As a general trend, also the quarks can easily be transformed
within one generation; that is analogous to the
conservation of the generation specific lepton numbers. 
This encompasses for instance the $\beta$-decay, 
${\rm n} \sim (udd) \to {\rm p} \sim (uud) + $ leptons.

However, transformations
between different generations happen as well: for instance,
the strangeness of a hadron changes when an $s$
quark decays into the much lighter quarks $u$ and $d$.
An example is the decay of the baryon $\Lambda^{0}$ 
into a nucleon and a pion,
\begin{eqnarray*}
\Lambda^{0} \to p + \pi^{-} & {\rm or } & \Lambda^{0} \to n + \pi^{0}\\
(uds) \to (uud) + (\bar u d) & & (uds) \to (udd) + 
(\bar u u - \bar d d)/\sqrt{2} \ ,
\end{eqnarray*}
(the lower line indicates the valence quark contents of the
hadrons involved). Based on the heavy $\Lambda^{0}$-mass of $1.1~{\rm GeV}$,
one could expect this decay to happen within $O(10^{-23})$ s, 
but since it proceeds only by the weak interaction it takes as long as
$2.6 \cdot 10^{-10}$ s.

The evolution is driven by the Hamiltonian, and from examples
like these strangeness changing decays we infer that the 
upper, or the lower, doublet partners
are {\em not} (exactly) its eigenstates.
Hence we have to distinguish the mass eigenstates 
$(u,c,t)$, or $(d,s,b)$, 
from the slightly different eigenstates of the 
weak interaction, $(u',c',t')$ or $(d',s',b')$, respectively. 

At this point, we recall that Dirac's 4-component spinor $\Psi$
actually describes a left-handed and a right-handed fermion;
the corresponding spinors are obtained by chiral projection
$\psi_{L,R} = \frac{1}{2} (1 \mp \gamma_{5}) \Psi$, cf.\ footnote
\ref{fnchiral}. The kinetic term in the Lagrangian keeps them apart, 
but the mass term involves both,
$m \bar \Psi \Psi = m (\bar \psi_{L} \psi_{R} + \bar \psi_{R} \psi_{L})$,
so $m>0$ breaks the chiral symmetry.

In terms of upper and lower quark doublet
components, the mass term takes the form
\be
- {\cal L}_{\rm quark~masses} = (\bar d'_{L} , \bar s'_{L},\bar b'_{L})
M_{d} \left( \begin{array}{c} d'_{R} \\ s'_{R} \\ b'_{R} \end{array} 
\right) + 
(\bar u'_{L} , \bar c'_{L},\bar t'_{L})
M_{u} \left( \begin{array}{c} u'_{R} \\ c'_{R} \\ t'_{R} \end{array} 
\right) \ .
\ee
A transformation to the mass base diagonalizes the matrices
$M_{d}$ and $M_{u}$, 
$U_{d;L}^{\dagger} M_{d} U_{d;R} = {\rm diag} (m_{d}, m_{s}, m_{b})$,
$U_{u;L}^{\dagger} M_{u} U_{u;R} = {\rm diag} (m_{u}, m_{c}, m_{t})$.
Thus the weak interaction eigenstates and the mass eigenstates
are related by unitary transformations,
\be
\left( \begin{array}{c} u' \\ c' \\ t' \end{array} \right)_{L,R}
= U_{u;L,R} \left( \begin{array}{c} u \\ c \\ t \end{array} \right)_{L,R}  
\ , \quad
\left( \begin{array}{c} d' \\ s' \\ b' \end{array} \right)_{L,R}
= U_{d;L,R} \left( \begin{array}{c} d \\ s \\ b \end{array} \right)_{L,R} \ ,
\ee
$U_{u;L,R}, \, U_{d;L,R} \in U(3)$.
The Standard Model describes the flavor changing due to the 
weak interaction by charged currents $J_{\mu}^{\pm}$,  such as
\be \label{jplusq}
J_{\mu}^{+} = (\bar u', \bar c', \bar t')_{L} \gamma_{\mu} 
\left( \begin{array}{c} d' \\ s' \\ b' \end{array} \right)_{L}
= (\bar u, \bar c, \bar t)_{L} \gamma_{\mu} 
\underbrace{U_{u;L}^{\dagger} U_{d;L}}_{V \in U(3)}
\left( \begin{array}{c} d \\ s \\ b \end{array} \right)_{L} \ .
\ee
Hence flavor changes are parameterized by a unitary matrix
$V$, known as the {\em Cabbibo-Kobayashi-Maskawa (CKM) matrix.}

For $N_{g}$ fermion generations it would be a matrix $V \in U(N_{g})$,
with $N_{g}^{2}$ real parameters. However, the diagonalization still
works if we vary any diagonal phase factor in $U_{u;L}$ and $U_{d;L}$,
so if we count the physical parameters, we should subtract these
$2N_{g}$ phases. On the other hand, one common phase in $U_{u;L}$ and 
$U_{d;L}$ leaves $V$ invariant, so that phase cannot be subtracted.
We end up with
$$
N_{g}^{2} - (2N_{g} -1) = (N_{g}-1)^{2}
$$
physical mixing parameters. 

This formula obviously works for one generation (nothing 
to be mixed). For $N_{g}=2$ there is only one rotation angle,
hence an SO(2) matrix is sufficient \cite{Cabbibo}; 
this is the Cabbibo angle, $\theta_{c} \approx 13^{\circ}$. 
For $N_{g}=3$ we obtain the 3 rotation angles ({\it e.g.}\ the Euler
angles) plus one complex phase. Kobayashi and Maskawa noticed that
this phase {\em breaks CP symmetry} \cite{KM} (if it doesn't vanish), 
so the aforementioned CP violation does naturally emerge in the
Standard Model with $N_{g} \geq 3$ generations.

The CKM matrix is well explored now by numerous experiments --- 
its unitarity was a theoretical prediction, which is compatible 
with the data. This is another argument why more than 3 fermion 
generations seem unlikely.
Actually $V$ is quite close to a unit matrix, with diagonal
elements $|V_{ii}| > 0.97$. Hence the off-diagonal elements, 
which enable the generation changes, are suppressed, but
the complex phase is clearly non-zero.

\subsection{$\dots$ and how about the leptons?}

The way the Standard Model was traditionally formulated, it does 
not include right-handed neutrinos (as we mentioned before), 
and all neutrino masses vanish. 
Still, there are flavor changing lepton currents,
in analogy to the quark current (\ref{jplusq}),
\be  \label{fermicur}
j_{\mu}^{+} = (\bar \nu_{\rm e}', \bar \nu_{\mu}', \bar \nu_{\tau}')_{L} 
\gamma_{\mu} \left( \begin{array}{c} e' \\ \mu' \\ \tau' \end{array} 
\right)_{L} = (\bar \nu_{\rm e}, \bar \nu_{\mu}, \bar \nu_{\tau})_{L}
\gamma_{\mu} U_{n;L}^{\dagger} U_{e;L}
\left( \begin{array}{c} e^{-} \\ \mu^{-} \\ \tau^{-} 
\end{array} \right)_{L}  \ .
\ee
However, in this case the choice of the matrix $U_{n;L}$ is completely
free --- if all neutrino masses vanish, there is no condition for
the diagonalization of their mass matrix. In particular we are free
to choose $U_{n;L} = U_{e;L}$, so the matrix, which would correspond
to the CKM matrix, can be set to $1 \!\! 1$. This shows that no physical
mixing effects --- analogous to the quark sector --- can be expected,
{\em in this original form of the Standard Model.}

We can turn this statement the other way round: if a transmutation of 
$\nu$-flavors is observed,
we can conclude that also for neutrinos the flavor and mass 
eigenstates differ, and therefore they cannot be all massless.
We now know that this is Nature's choice, as we are going to
review next.  

\section{Neutrino oscillation: \\ a chameleon-like metamorphosis}

In 1957 Pontecorvo formulated a first idea that neutrinos could
somehow transform into each other \cite{Pont57}. This early suggestion 
was an oscillation between neutrino and anti-neutrino, 
$\nu \leftrightarrow \bar \nu$, which would violate
the conservation of the lepton number $L$.
In 1962, the year when the neutrino $\nu_{\mu}$ was discovered,
Ziro Maki, Masami Nakagawa and Shoichi Sakata at Nagoya University 
(Japan) considered the possibility
of massive neutrinos, and suggested that their mass eigenstates 
could be superpositions of $\nu_{\rm e}$ and $\nu_{\mu}$ \cite{MNS}. 
In 1968 it was again Pontecorvo who elaborated a full-fledge 
theory for this scenario \cite{Pont68}, and for the resulting
$\nu_{\rm e} \leftrightarrow \nu_{\mu}$ oscillation,
which changes the generation-specific lepton numbers
$L_{\rm e}$ and $L_{\mu}$, but not $L$. 

This 2-flavor setting is convenient for illustration:
we denote the mass eigenstates as $\nu_{1}$, $\nu_{2}$. As we
saw in the discussion of the CKM quark mixing matrix, this case only
involves one physical mixing parameter, namely the rotation angle 
of an SO(2) matrix,
$$
\left( \begin{array}{c} \nu_{\rm e} \\ \nu_{\mu} \end{array} \right)
= \left( \begin{array}{cc} ~  \ \cos \theta & \sin \theta \\
-\sin \theta & \cos \theta \end{array} \right)
\left( \begin{array}{c} \nu_{1} \\ \nu_{2} \end{array} \right) \ .
$$
Let us assume a plane wave dynamics for the mass eigenstates,
which we write as {\em kets} (in Dirac's notation),
$$
| \nu_{i} (t) \ra = \exp (- \ri (E_{i} \, t-\vec p_{i} \cdot \vec r)) \
| \nu_{i} (0)\ra \ , \quad (i=1,2) \ .
$$
The distance that the neutrino  has travelled --- after its start
at time $t=0$ --- is (in natural units) $L \simeq t\,$; 
the mass is so small that it is ultra-relativistic even
at modest energy. This also implies 
$m_{i} \ll |\vec p_{i}| =p_{i} \approx E_{i}$, and we obtain 
$$ E_{i} - p_{i} = \sqrt{p_{i}^{2} + m_{i}^{2}} - p_{i}
\approx m_{i}^{2} /(2 p_{i}) \approx  m_{i}^{2} /(2 E_{i}) \ ,
$$
which simplifies the propagation to
$$
| \nu_{i} (t) \ra = \exp (- \ri m_{i}^{2} L/(2E_{i})) \ | \nu_{i} (0) \ra \ .
$$
In the framework of this approximation, an initial
state $| \nu_{\rm e}\ra $ is converted into $| \nu_{\rm \mu} \ra $ 
(or vice versa), after flight distance $L$, with probability
(for a derivation, see {\it e.g.}\ Ref.\ \cite{Pilar})
\be  \label{transprob}
P_{{\rm e} \leftrightarrow \mu} = | \la \nu_{\mu} | \nu_{\rm e} \ra |^{2} 
= \sin^{2}(2\theta) \sin^{2} \Big( \frac{\Delta m_{12}^{2} L}{4 E} \Big) \ , 
\quad \Delta m_{12}^{2} = m_{2}^{2} - m_{1}^{2} \ .
\ee
Intuitively, the initial state $| \nu_{\rm e} \ra$ consists of a
peculiar superposition of $| \nu_{1} \ra$ and $| \nu_{2} \ra$,
but these components propagate with different speed. Therefore
the composition changes to new states, which mix $| \nu_{\rm e} \ra$
and $| \nu_{\mu} \ra$.\\

It is straightforward to extend this approach to the case of
3 flavors and 3 mass eigenstates $| \nu_{i} \ra$,
\be  \label{PMNSmatrix}
\left( \begin{array}{c} \nu_{\rm e} \\ \nu_{\mu} \\  \nu_{\tau} \end{array}
\right) = U_{\rm PMNS} \left( \begin{array}{c} 
\nu_{1} \\ \nu_{2} \\  \nu_{3} \end{array} \right) \ , \quad
U_{\rm PMNS} \in U(3) \ ,
\ee
where $U_{\rm PMNS}$ is the {\em Pontecorvo-Maki-Nakagawa-Sakata (PMNS) 
matrix.} As we saw in the case of the CKM matrix, there are now 3 
mixing angles plus one complex phase, which implies additional
CP symmetry breaking, now in the lepton sector.

In this case, the oscillation probability is
$\propto \sin^{2} (\Delta m^{2}_{ij} L/(4E))$, so we can determine
$|\Delta m^{2}_{12}|$, $|\Delta m^{2}_{23}|$ and $|\Delta m^{2}_{13}|$
(they are not independent, hence one can focus on two of them).

Experiments are built with a given average neutrino energy $E$ and a
fixed baseline $L$. If two $|\Delta m^{2}_{ij}|$ are sufficiently 
different, an appropriate ratio $L/E$ selects to which one the 
experiment is most sensitive. Initially this separability was
uncertain, but fortunately for the experimentalists
it turned out that $|\Delta m^{2}_{12}| \approx 30 \ |\Delta m^{2}_{23}|$.
The former (latter) was crucial for the observation of solar
(atmospheric) neutrinos, see below.

So this can be tested experimentally, but
in practice it is a delicate task: many attempts
to probe this behavior ended up with results that were not fully
conclusive. This changed at the dawn of the new millennium,
with the experiments that were awarded the 2015 Nobel Prize.

\subsection{Atmospheric neutrinos viewed \\ by Super-Kamiokande}

In 1996 the experiment {\em Super-Kamiokande} was launched, as an
extension of the previous Kamiokande. It is located in the Mozumi 
zinc mine, near the town Kamioka (now part of Hida) in central 
Japan, about 1000 m underground. Such locations deep underground are 
standard for neutrino experiments (and also for Dark Matter search),
because of the shielding from the background radiation, which is a 
major challenge for the experimentalists.

Super-Kamiokande used 50\,000 t of water as a Cherenkov detector.
It focused on {\em atmospheric neutrinos,} which we briefly mentioned
in Section 3: high energy cosmic rays hit our atmosphere and
generate a shower of secondary particles, in particular
light mesons (pions and kaons), which subsequently decay into
leptons, including neutrinos. Examples are the charged pion decays,
$$
\pi^{+} \to \mu^{+} + \nu_{\mu} , 
\ \mu^{+} \to {\rm e}^{+} + \nu_{\rm e} + \bar \nu_{\mu} 
\quad {\rm or} \quad
\pi^{-} \to \mu^{-} + \bar \nu_{\mu} , 
\ \mu^{-} \to {\rm e}^{-} + \bar \nu_{\rm e} + \nu_{\mu} \ ,
$$
{\it i.e.}\ successions of the decays (\ref{pitomu}) and (\ref{mudecay}).
The flux of cosmic rays is well-known, so also the resulting
neutrino flux could be predicted: the ratio between
the number of $\mu$-(anti-)neutrinos and e-(anti-)neutrinos should be 
about 2:1, as in our example. Cosmic rays arrive isotropically,
and --- as we mentioned in Section 3 --- crossing
the Earth reduces the neutrino flux only by a negligible
fraction of $O(10^{-18})$. Does this mean that the neutrino flux 
observed in the Mozumi mine is isotropic as well?

Super-Kamiokande monitored neutrino reactions, which involve
charged currents and emit ${\rm e}^{\pm}$ or $\mu^{\pm}$;
examples are given in scheme (\ref{CC}).
This causes water Cherenkov radiation, which indicates the neutrino
direction and energy; the high energies --- up to several
GeV --- distinguish them from the background neutrinos.
The profile of the Cherenkov cone further reveals whether it was
triggered by an ${\rm e}^{\pm}$ or by a $\mu^{\pm}$, and therefore
whether its origin was an atmospheric e- or $\mu$-neutrino
(though $\nu$ and $\bar \nu$ could not be distinguished).

For the $\nu_{\rm e}$ and $\bar \nu_{\rm e}$ flux, the prediction
was well confirmed, and its isotropy too. This was {\em not}
the case for the  $\nu_{\mu}$ and $\bar \nu_{\mu}$ flux:
here part of the expected neutrinos were missing, and
the flux from above was significantly larger than the one
from below (after passing through the Earth). This was announced 
in 1998, after two years of operation, based on 5000 neutrino 
signals \cite{SuperK}.

In light of this section, the explanation 
is clear: part of the missing $\mu$-neutrinos were transformed into
$\tau$-neutrinos! This oscillation takes a while, this is why
it happens mostly along the extended path across the Earth.
The precise angular distribution reveals the oscillation
rate as a function of the travelling distance $L$, 
divided by the $\nu_{\mu}$ energy $E$.
This determines the difference 
$| \Delta m_{23}^{2} | = | m_{3}^{2} - m_{2}^{2}| \approx 2.4 
\cdot 10^{-3} ~{\rm eV}^{2}$. That has been confirmed later by 
experiments with accelerator neutrinos, which attain $O(1)~{\rm GeV}$. 
\begin{figure}
\vspace*{-2mm}
\begin{center}
\includegraphics[angle=0,width=.53\linewidth]{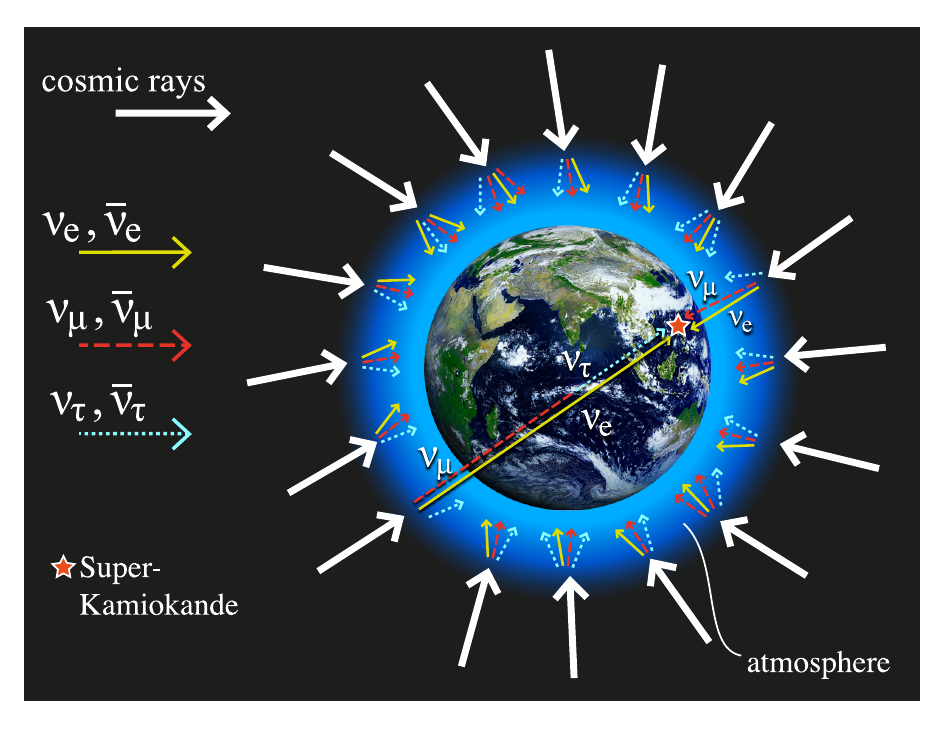}
\hspace*{2mm}
\includegraphics[angle=0,width=.42\linewidth]{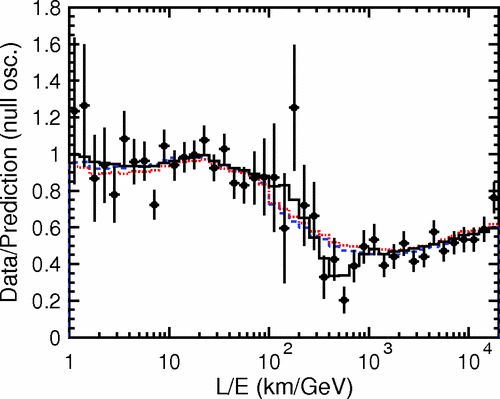}
\end{center}
\vspace*{-5mm}
\caption{\it{On the left: Illustration of the Super-Kamiokande experiment
on atmospheric neutrinos. Cosmic rays generate air showers of
secondary particles, including neutrinos. The e-neutrino flux
arrives as predicted, but for a long path part of the $\mu$-neutrinos
are converted into $\tau$-neutrinos. 
On the right: The atmospheric $\nu_{\mu}$ plus $\bar \nu_{\mu}$ 
flux, as observed by Super-Kamiokande, as a function of the 
travelling distance $L$ divided by the neutrino energy $E$.
The vertical axis is the ratio between measured flux and
the prediction without neutrino oscillation.}}
\vspace*{-2mm}
\end{figure} 

\subsection{The solar neutrino puzzle and its solution by SNO}

Almost all our activities are driven by solar energy.
For $4.5 \cdot 10^{9}$ years the Sun has been shining with
a luminosity of $3.8 \cdot 10^{26} ~ {\rm W}$, and it is expected 
to continue doing so for another  $4.5 \cdot 10^{9}$ years.
Until the 19th century the origin of all this energy
seemed mysterious: a chemical process was assumed, but estimates
showed that the Sun could only burn for 6000 years, even under
the ``most optimistic assumption'' that it consisted of coal.

In the 20th century {\em nuclear fusion} was identified as the energy 
source of the Sun, in particular the ``pp chain reaction'', 
which amounts to
$$
{\rm p} + {\rm p} + {\rm p} + {\rm p} \to \ \dots \ \to 
\ ^{4}{\rm He} + 2 {\rm e}^{+} + 2 \nu_{\rm e} \ .
$$
If we divide the solar luminosity by the energy, which is
released by this chain reaction ($26.7~{\rm MeV}$), we obtain 
the fusion rate, as well as an estimate for the $\nu_{\rm e}$ 
production ($\approx 2 \cdot 10^{38}~ {\rm s}^{-1}$).
In addition there are sub-dominant processes,
which emit electron neutrinos of higher energies.

The entire spectrum ranges from about $E_{\nu_{\rm e}} \approx (0.1 \dots 10) 
~ {\rm MeV}$. Since the 1960s the $\nu_{\rm e}$-flux 
arriving at the Earth was quite well predicted \cite{Schwarzschild}, 
and also measured --- first in the 
Homestake gold mine in South Dakota \cite{Homestake} --- but the data 
confirmed only about 1/3 of the expected flux. 
This {\em solar neutrino puzzle} (see {\it e.g.}\ Ref.\ \cite{solarnu})
persisted for more than 30 years. 

Various solutions were discussed, such as corrections to the
solar model, but the latter was constantly improved, in particular
by John Bahcall and collaborators, which led to the Standard Solar 
Model \cite{Bahcall}. This model was refined to a point that made it 
truly difficult to still raise objections which could reduce the 
$\nu_{\rm e}$-flux that much. Another explanation, which had been discussed
for decades, was finally confirmed in  2001: the solution to this
puzzle is {\em neutrino oscillation} --- this scenario had been suggested 
first by V.N.\ Gribov and B.\ Pontecorvo in 1969 \cite{GribPont}.

The breakthrough was due to the {\em Sudbury Neutrino Observatory (SNO)}
in Ontario, Canada, 2000 m underground \cite{SNO}. 
In its crucial experiment, 9500 photomultipliers 
monitored a sphere with 6 m radius, which contained 1000 t
of {\em heavy water,} D$_{2}$O (compared to ordinary water, 
a neutron is added to each proton, thus forming deuterium, 
D$\, \sim \, $(np)).
This offered several options for the detection of neutrino events:
\begin{itemize}

\item The variant of the $\beta$-decay shown in Figure 3, with an 
incoming $\nu_{\rm e}$ and an outgoing electron; this
measures exclusively the $\nu_{\rm e}$ flux.

\item A deuterium dissolution, 
${\rm D} + \nu_{x} \to {\rm n} + {\rm p} + \nu_{x}\ , 
\ x \in \{ {\rm e},\mu , \tau \} \, $. 
That process measures the total neutrino flux without flavor distinction, 
{\it i.e.}\ the sum of $\nu_{\rm e}$, $\nu_{\mu}$ and $\nu_{\tau}$ neutrinos.

\item Elastic $\nu_{x} \, {\rm e}^{-}$ scattering enables a good
identification of the direction, which affirmed that the observed
neutrino flux indeed originates from the Sun. (Only for $\nu_{\rm e}$ 
the scattered particles can also be exchanged.)

\end{itemize}

The total flux is well compatible 
with the prediction by the Standard Solar Model. On the other hand, 
this model predicts solely $\nu_{\rm e}$-production, but the first process
accounts for only $\approx 1/3$ of the expected $\nu_{\rm e}$-flux, 
in agreement with earlier experiments. 
Taken together, these results imply
that 2/3 of the solar $\nu_{\rm e}$ have been transformed into other 
flavors before they reach us. 

If neutrinos can oscillate, we can expect all flavors
to be equally frequent after a long path, like the 
$1.5 \cdot 10^{11}~{\rm m}$ that separate us from the Sun,
which yields a $\nu_{\rm e}$ survival probability of $1/3$.
Moreover, neutrino oscillation takes place already inside the Sun, 
before the neutrinos even leave it, enhanced by the medium \cite{medium}.

This is the ultimate demonstration that neutrino oscillation {\em is} 
the solution to the long-standing solar neutrino puzzle, as Gribov
and Pontecorvo had conjectured.

\begin{figure}
\begin{center}
\includegraphics[angle=0,width=.53\linewidth]{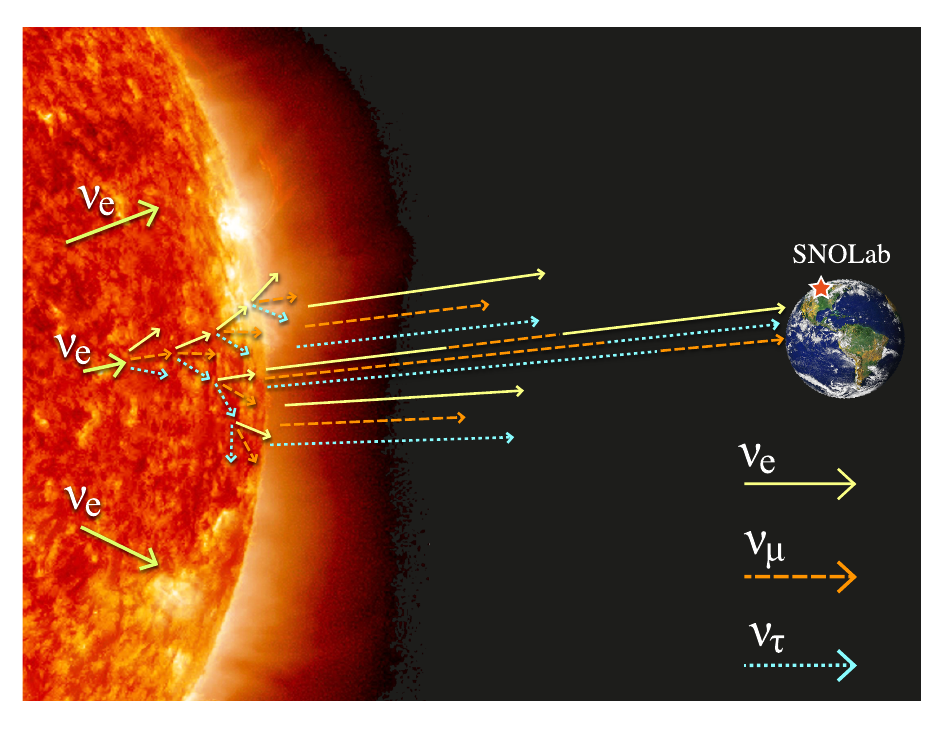}
\hspace*{2mm}
\includegraphics[angle=0,width=.42\linewidth]{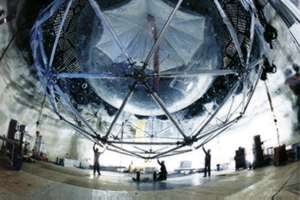}
\end{center}
\vspace*{-5mm}
\caption{\it{The Standard Solar Model predicts 
the generation of numerous electron neutrinos $\nu_{\rm e}$
inside the Sun ($\approx 2 \cdot 10^{38} \ {\rm s}^{-1}$), 
such that a flux of $\approx 6 \cdot 10^{10}$ 
$\nu_{\rm e}$/(cm$^{2} \cdot$s)} was expected at the Earth. Only 1/3
of them arrive as $\nu_{\rm e}$, the rest is transmuted into
$\nu_{\mu}$ or $\nu_{\tau}$ by means of neutrino oscillation,
as illustrated on the left. This was 
demonstrated by the SNO Laboratory, which used a spherical 
detector filled with 1000 t of heavy water, shown on the right.}
\vspace*{-2mm}
\end{figure} 

\section{Status today: PMNS matrix and open questions}

Meanwhile a host of experiments confirmed these observations
on atmospheric and solar neutrinos: some detected reactor
neutrinos at distances of $O(100) ~{\rm km}$, confirming the
atmospheric $\nu_{\mu} \leftrightarrow \nu_{\tau}$ oscillation,
while accelerator neutrinos are consistent with the solar
$\nu_{\rm e} \leftrightarrow \nu_{\mu}, \, \nu_{\tau}$ transmutation.
By global fits, the absolute values of the PMNS matrix elements 
in eq.\ (\ref{PMNSmatrix}) are quite well determined, 
$$
\left( \begin{array}{ccc} 
|U_{{\rm e} 1}| & |U_{{\rm e} 2}| & |U_{{\rm e} 3}| \\
|U_{\mu 1}| & |U_{\mu 2}| & |U_{\mu 3}| \\ 
|U_{\tau 1}| & |U_{\tau 2}| & |U_{\tau 3}| \end{array} \right)
= \left( \begin{array}{ccc} 
0.82(2) & 0.55(3) & 0.15(1) \\
0.37(15) & 0.57(13) & 0.70(9) \\
0.39(14) & 0.59(12) & 0.68(9) \end{array} \right) \ .
$$
The reduction of the uncertainties is in progress.

The dark horse is the {\em complex phase:} it depends on the parameterization 
how it occurs in this matrix, but the physically interesting aspect 
of a leptonic CP violation is still highly uncertain.

This could be relevant for the famous puzzle about the
{\em matter--anti-matter imbalance} in the Universe: the Big Bang should 
have generated the same amount of both, so how comes that today there is
an enormous dominance of matter? One of the three conditions for a possible
explanation (formulated by Andrei Sakharov in 1967 \cite{Sakharov}) is 
CP violation. 
We have mentioned that this was indeed observed in weak decays, and that
the complex phase in the CKM matrix breaks CP invariance, but this violation
is not sufficient to account for the striking matter--anti-matter 
imbalance. In this regard, an additional CP violation in the lepton sector
could be helpful.

Regarding the neutrino masses, 
it seems natural to assume that the flavors follow the same 
hierarchy as the charged leptons, $m_{1} < m_{2} < m_{3}$.
However, since the neutrino oscillation between any two flavors in 
vacuum only determines $| \Delta m^{2}|$, cf.\ eq.\ (\ref{transprob}),
an ``inverse hierarchy'' with $m_{3} < m_{1} < m_{2}$ cannot be ruled 
out either (so far only $m_{1} < m_{2}$ is considered safe, based on 
processes inside the Sun).

In any case, we see that this mixing matrix is much more animated
than its counterpart in the quark sector;
neutrinos mix strongly! The element
with the least absolute value is $U_{{\rm e}3}$; for quite a while it
seemed to be compatible with 0, and people invented theories
to explain its possible vanishing --- until 2012, when the Chinese
reactor experiment Daya Bay, as well as RENO in South Korea and
Double Chooz in France, showed that it differs from 0, with more than
$5 \sigma$ significance (here the baseline was just $O(1)$ km) \cite{Daya}.

Generally, the attempts to search for a systematic ``texture''
in the PMNS matrix were not that fruitful --- it seems that we just
have to accept the values of its physical parameters as 
experimental input.

Moreover, this still leaves the question open how large the 
neutrino masses really are ---  
the PMNS matrix only contains information about
their mass squared {\em differences.} The masses themselves are
even more difficult to determine, and alternative techniques are 
required: one approach is the study of the $\beta$-decay to an 
extreme precision --- in particular the electron spectrum near
the endpoint is slightly sensitive to the neutrino mass. Such a study
is ongoing in the KArlsruhe TRItium Neutrino (KATRIN) experiment in 
Germany \cite{KATRIN}, which has the potential to improve the current
bound of $m_{\nu_{\bar {\rm e}}} < 2.3 ~{\rm eV}$ (by the experiments
Mainz in Mainz \cite{Mainz} and Troitsk in Russia \cite{Troitsk}) 
by an order of magnitude.

There are also cosmological estimates and bounds for the
neutrino masses (an overview is given in Ref.\ \cite{Valle}), 
though they necessarily involve some model dependence. 
In any case, the absolute values will be relevant 
for cosmology. Even if the neutrino masses are tiny,
their sum --- all over the Universe one estimates $O(10^{89})$ 
neutrinos --- could well be powerful: for instance, the exact masses 
could, along with the amount of Dark Matter,
be crucial for our long-term future, regarding the question if the 
Universe will keep on expanding for ever, 
or if it will end in a Big Crunch --- let's see $\dots$

\begin{figure}
\begin{center}
\includegraphics[angle=0,width=.36\linewidth]{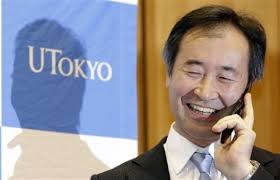}
\includegraphics[angle=0,width=.36\linewidth]{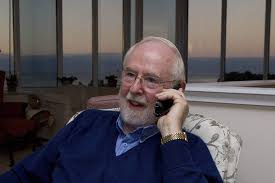}
\includegraphics[angle=0,width=.24\linewidth]{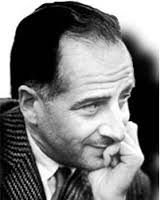}
\end{center}
\caption{\it{Left and center: Takaaki Kajita and Arthur McDonald,
Nobel Prize laureates 2015. On the right: Bruno Pontecorvo.
\newline
$\bullet$ Kajita (born 1959) studied at Saitama University and completed his 
Ph.D.\ 1986 at Tokyo University, where he later worked in the
Institute for Cosmic Radiation Research. He led the group at
Super Kamiokande, which found evidence for the oscillation of atmospheric
neutrinos. In 1999 he became director of the Research
Center for Cosmic Neutrinos in Tokyo.
\newline
$\bullet$ McDonald (born 1943) studied at Dalhousie University 
(Halifax, Canada) and 
did his Ph.D.\ at the California Institute of Technology. He worked from
1970 to 1982 at the Chalk River Laboratories near Ottawa, from 1982 
to 1989 at Princeton University, then he became director of the Sudbury
Neutrino Observatory (SNO), which solved the solar neutrino puzzle.
\newline
$\bullet$ If he were still alive, then Pontecorvo (1913-1993) should 
be another 2015 Nobel Prize winner, as the leading theorist involved.
He worked in Rome with Enrico Fermi, 
and later in Paris, Montreal and Liverpool.
In 1950 he moved to the Joint Institute for Nuclear Research (JINR)
in Dubna (near Moscow), where he elaborated the theory of neutrino 
oscillation. On this basis, he and Vladimir Gribov predicted in 1969 
the correct solution to the solar neutrino puzzle.}}
\end{figure}
\ \\

\noindent
{\bf Acknowledgements} \ \  
This work was supported by UNAM-PAPIIT through grants IN107915,
IN112213 and IB100413.

\appendix

\section{Neutrino masses are still puzzling}

In the traditional form of the Standard Model, the first
fermion generation contains the following leptons and quarks,
$$
\left( \begin{array}{c} \nu_{{\rm e};L} \\ {\rm e}_{L} \end{array} 
\right) , \ {\rm e}_{R} \ ,
\quad 
\left( \begin{array}{c} u_{L} \\ d_{L} \end{array} \right) , 
\ u_{R}, \ d_{R} \ ,
$$
where we now keep track of left- and right-handed fermions separately. 
For instance the term for the electron mass $m_{\rm e}$ takes the form
$m_{e} (\bar {\rm e}_{R} {\rm e}_{L} + \bar {\rm e}_{L} {\rm e}_{R})$.
However, this explicit mass term must not appear in the Lagrangian:
${\rm e}_{L}$ and ${\rm e}_{R}$ couple differently to the
electroweak gauge fields, so this term would break gauge invariance.

Instead the Higgs field 
$$
\Phi = \left( \begin{array}{c} \phi_{+} \\ \phi_{0} \end{array} 
\right) \in \CC^{2}
$$
comes to the rescue and endows the gauge invariant Yukawa term
$$
- {\cal L}_{\rm Yukawa} = f_{e} \ \Big[ \, \bar {\rm e}_{R} \ \Phi^{\dagger} \cdot 
\left( \begin{array}{c} \nu_{{\rm e};L} \\ {\rm e}_{L} \end{array} \right) 
+ ( \bar \nu_{{\rm e};L}, \bar {\rm e}_{L}) \cdot \Phi \ {\rm e}_{R} \, \Big] \ ,
$$
where $f_{e}$ is a (dimensionless) Yukawa coupling.
The Higgs potential arranges for spontaneous symmetry breaking.
If the Higgs field takes the classical ground state configuration
$$
\Phi_{0} = \left( \begin{array}{c} 0 \\ v \end{array} \right) ,
\ v \simeq 246 ~ {\rm GeV} \ \Rightarrow \
{\cal L}_{\rm Yukawa} = - f_{e} v \, [ \bar {\rm e}_{R} {\rm e}_{L}
+ \bar {\rm e}_{L} {\rm e}_{R} ] , \  m_{e} = f_{e} v \ ,
$$
while the neutrino remains massless.\\

The analogous term for the quark doublet (with a Yukawa coupling
$f_{d}$) leads to the $d$-quark mass $m_{d}= f_{d} v$.
But how do we give mass to the $u$-quark? One could introduce an 
additional Higgs field, but the Standard Model is economic and
recycles $\Phi$: another quark Yukawa term is added, with 
$\tilde \Phi = \left( \begin{array}{c} \!\! - \phi_{0}^{*} \\ 
\!\! ~~\phi_{+}^{*} \end{array} \right)$ instead of $\Phi$, and we 
obtain $m_{u}= -f_{u} v$ ($f_{u} < 0$ is allowed).

If we want to construct a {\em neutrino mass,} we can do exactly
the same, {\em if} we add a right-handed neutrino, $\nu_{{\rm e};R}$.
It turns out that $\nu_{{\rm e};R}$ is ``sterile''; it does not have 
any charge, so it does not couple to any gauge field.
It could have hidden from our detectors, and it is a Dark Matter 
candidate.

One often hears the statement that the neutrino mass is
``beyond the Standard Model''. While this is ultimately a matter
of semantics, we would like to emphasize that neutrino masses 
can be constructed in the same way as it is done
for the $u,\, c,$ and $t$-quark, so this does not necessarily 
require a conceptual extension of the Standard Model. 

Alternative approaches do speculate about conceptual novelties,
like a dimension 5 mass term,\footnote{A term of this kind is $\propto 
\Big[ (\bar \nu_{L}, \bar {\rm e}_{L}) \cdot \tilde \Phi \Big] \, 
\Big[ \tilde \Phi^{\dagger} \cdot 
\left( \begin{array}{c} \nu_{L} \\ {\rm e}_{L} \end{array} \right) \Big]$,
which is not renomalizable, but it does not require any sterile neutrino.}
or even scenarios in higher space-time dimensions,
but we are not going to discuss them.\\

We just add that the presence of $\nu_{R}$ opens the door 
to new scenarios (we do not specify the generation anymore). 
In general, the C transformation (charge conjugation) 
of a fermion field $\Psi$ reads
$$
{\rm C}~:~\Psi \to \Psi^{C} = C \bar \Psi^{T} \ ,
$$
where $T$ means ``transposed'', and $C$ is a matrix that fulfills 
suitable conditions. Therefore the {\em Majorana spinors}
$$
\nu_{1}^{\rm M} = \nu_{R} + C \bar \nu_{R}^{T} \doteq \nu_{R} + \nu_{L}^{C} \ ,
\quad \nu_{2}^{\rm M} = \nu_{L} + C \bar \nu_{L}^{T} \doteq \nu_{L} +\nu_{R}^{C}
$$
are C-invariant; each of them represents a {\em Majorana neutrino,} 
which is its own anti-particle. In one generation we obtain
one Majorana neutrino with the chirality components $\nu_{R}$ 
and $\bar \nu_{L}$, and another one with $\nu_{L}$ and $\bar \nu_{R}$.

This construction yields real, {\it i.e.}\ neutral spinor fields.
In Dirac's and Weyl's original approaches, the $\gamma$-matrices 
are chosen such that the Dirac operator $(\ri \gamma^{\mu} \partial_{\mu} - m)$
contains complex elements, which was considered as an argument
that fermions should have some charge, and the corresponding
operators generate distinct particles and anti-particles.

However, in the 1930s Ettore Majorana found a way to fulfill the 
conditions of the Dirac algebra
($\gamma^{\mu} \gamma^{\nu} + \gamma^{\nu} \gamma^{\mu} = 2 g^{\mu \nu}$) 
with purely imaginary $\gamma$-matrices, such that the
Dirac operator becomes entirely real,
which disproved this argument, and showed that 
neutral fermions are another option.\footnote{Majorana did not publish 
the work with this insight himself, but he told Fermi about it, and allowed 
him to do so in his name. This paper appeared in 1937 \cite{Majorana}, 
one year before Majorana mysteriously disappeared. 
For a semi-popular review, see Ref.\ \cite{Wilczek}.}

In fact, it is conceivable that the neutrinos are Majorana 
particles, and not ``Dirac neutrinos'' as we assumed in the main 
part of this article. Then the counting of the physical
parameters in the mixing matrix has to be reconsidered:
roughly speaking, we argued before that the $U(3)$ matrix 
in eq.\ (\ref{fermicur}) has 9 parameters, but --- with massive
neutrinos --- each fermion field in the current $j^{+}_{\mu}$ 
can absorb one phase (but one common phase cancels), so we 
are left with $9 - (6 -1) = 4$ physical parameters.
If we insert Majorana neutrinos instead, these three fields cannot
absorb any phase, and there is no common phase either. So in that
case there are $9-3 = 6$ physical parameters, which include 
{\em 3 complex phases.}

For Majorana fermions, an explicit mass term
$$
{\cal L}_{\rm Majorana~mass} = - \frac{\cal M}{2} \bar \nu^{\rm M} \nu^{\rm M}
$$
can be incorporated directly in the Lagrangian.
Then the theory contains another dimensional parameter, the
Majorana mass ${\cal M}$ (not related to the Higgs mechanism), 
in addition to $v$, without breaking gauge symmetry. 
It does, however, break the conservation of the total 
lepton number $L = L_{\rm e}+L_{\mu}+L_{\tau}$.\footnote{It also changes
the difference between baryon and lepton number, $B-L$. This is
the quantity, which is strictly conserved in the Standard Model.
There combined $B$ and $L$ anomalies are conceivable, but not observed.}
After the observation that neutrino oscillation violates the 
separate $L_{\rm e}$, $L_{\mu}$ and $L_{\tau}$ conservation, could 
it be that not even $L$ is on safe ground? 

Back in 1939, Wendell Furry pointed that a {\em neutrinoless
double $\beta$-decay} $2n \to 2 p + 2 {\rm e}^{-}$ 
would confirm this scenario \cite{Furry}; it changes $L \to L+2$.
This is a way how experiment could confirm that neutrinos are of
Majorana type, and --- by means of the decay rate --- also
explore their masses \cite{Rode}.
The ordinary double $\beta$-decay (with $2\bar \nu_{\rm e}$ emission) 
has been observed since 1987 \cite{doubleb}, but the hunt for its 
{\em neutrinoless} counterpart is still going on: some
events were reported, but the community is not convinced.\footnote{A
drama began in 2001, when part of the Heidelberg-Moscow Collaboration
claimed evidence for the decay 
$^{76}_{32}{\rm Ge} \to \ ^{76}_{34}{\rm Se} + 2 {\rm e}^{-}$, but it was 
refuted by other experts, including members of the same collaboration.}
The consensus so far is a lower bound of $\approx 2 \cdot 10^{25}$ 
years for the corresponding life time.\\

Last but not least, Majorana neutrinos enable the 
{\em seesaw mechanism,} which is popular as a possible explanation
why neutrinos are so light (a ``hierarchy problem'').
It was suggested by Peter Minkowski in 1977 \cite{Mink} and others;
here we illustrate its simplest form (``type 1'') in one generation.

We endow the Majorana spinor fields $\nu^{\rm M}_{1}$, $\nu^{\rm M}_{2}$
with a ``Dirac mass'' $\ri m$ (a coupling between components of
distinct Majorana fields with different chirality; for later convenience
we choose it imaginary), and a ``Majorana mass'' $M$
(it would be the Majorana mass of $\nu^{\rm M}_{1}$,
in the absence of $\nu^{\rm M}_{2}$),
$$
- {\cal L}_{\rm neutrino~masses} = \frac{1}{2} (\bar \nu_{L}, \bar \nu_{L}^{C})
\left( \begin{array}{cc} 0 & \ri m \\ \ri m & M \end{array} \right)
\left( \begin{array}{c} \nu_{R}^{C} \\ \nu_{R} \end{array} \right) 
+ {\rm Hermitian~conjugate} \ .
$$
Really physical are the Majorana masses for the eigenstates, 
{\it i.e.}\ the eigenvalues of this matrix.
In particular, for $M \gg m$ we obtain
$$
{\cal M}_{\rm large} \simeq M \ , \quad
{\cal M}_{\rm small} \simeq \frac{m^{2}}{M} \ll {\cal M}_{\rm large} \ .
$$
The more we amplify ${\cal M}_{\rm large}$ (by increasing $M$), 
the more we suppress ${\cal M}_{\rm small}$. This setting of 
injustice inspired the term ``seesaw mechanism''.

If we choose $m$ somewhat above 
the vacuum expectation value of the Higgs field, 
$v \lesssim m = O(1)~{\rm TeV}$, and insert a huge 
$M \approx 10^{24} \dots 10^{25} ~{\rm eV}$, we obtain a very light 
neutrino, with a realistic mass 
${\cal M}_{\rm small} \approx 0.1 \dots 1 ~{\rm eV}$.
In this scenario, ${\cal M}_{\rm large}$ has the magnitude of the energy,
where a Grand Unification of the electroweak and strong interactions is 
expected (``GUT scale'', somewhat below the Planck scale 
$\approx 10^{28}~{\rm eV}$), which many theorists find appealing.\\

Hence in some sense history is repetitive: as in 1930, there are
strong theoretical reasons for postulating a hypothetical particle,
now it is the sterile neutrino $\nu_{R}$. It is even more elusive 
than the known, weakly interacting neutrinos, but it could possibly
fix several short-coming in the traditional form of the Standard Model,
while preserving its elegant and successful concepts:
it provides a sensible Dark Matter candidate, and neutrino masses 
appear in an natural way. We can even explain why they are so light,
if we assume the seesaw mechanism. Then primordial heavy Majorana neutrinos
should have decayed in the very early Universe, generating slightly more
anti-leptons than leptons (``leptogenesis''). A cascade of further decays 
would generate an extreme excess of baryons over anti-baryons 
(``baryogenesis''), and thus the dominance of matter that we still
experience today.\footnote{The possible impact of sterile neutrinos
in astrophysics and cosmology is reviewed in Ref.\ \cite{nur}.}

So postulating $\nu_{R}$ (in 3 generations) seems to be a good deal,
but its experimental search is a tremendous issue: {\it e.g.}\ the
Chandra X-Ray Observatory searches for faint pulses from their
possible decay into lighter neutrinos, while the
Wilkinson Microwave Anisotropy Probe (WMAP) measures tiny fluctuations in
the Cosmic Microwave Background, which could indicate the likelihood
of 4 neutrino types. 
Clear evidence is still missing, although some hints of its existence 
may be hiding in a few anomalous experimental results \cite{WhitePap}. 
So the ghost hunt for the sterile neutrino $\nu_{R}$ is going on, and 
neutrino physics will continue to be exciting in the future.

\end{document}